# Algorithmic Accuracy as a Motivational Driver in Robot-Mediated Learning: A Comparative Study of Cross-Correlation and CNN-Based Sound Detection in an Interactive Quiz Game

Rezaul Tutul[1], Ilona Buchem[2] and Niels Pinkwart[1]

[1] Humboldt University of Berlin, Berlin, Germany
[2] Berlin University of Applied Science, Berlin, Germany
rtutul07@gmail.com

**Abstract.** The increasing adoption of social robots in education has created new opportunities for interactive and engaging learning experiences. While previous research has primarily focused on how robot appearance, behaviour, and pedagogical strategies influence learner motivation, comparatively little attention has been given to the impact of the robot's underlying perception algorithms. In competitive learning activities, inaccurate robot decisions may reduce students' perceptions of fairness and competence, ultimately affecting their motivation. This paper investigates whether the accuracy of sound detection algorithms influences student motivation during a robot-mediated quiz game. A Pepper humanoid robot hosted an interactive buzzer-based quiz in which two sound detection approaches, a Convolutional Neural Network (CNN) and a Cross-Correlation algorithm, were evaluated using a controlled between-subjects experiment involving 40 university students. Participants were equally assigned to a CNN group ($n = 20$) and a Cross-Correlation group ($n = 20$). Both groups completed the same quiz under identical conditions, differing only in the sound detection algorithm used for first-responder identification. Student motivation was assessed using the Intrinsic Motivation Inventory (IMI), while algorithm performance was evaluated through real-time detection accuracy. The results indicate that the Cross-Correlation approach achieved more reliable sound detection under classroom conditions and produced significantly higher scores across all IMI subscales, demonstrating greater student interest, perceived competence, effort, perceived choice, and lower perceived pressure (after reverse coding). These findings provide empirical support for the proposed Algorithmic Precision–Motivation Relationship (APMR) model, demonstrating that algorithmic accuracy is not merely an engineering performance metric but an important factor influencing learner motivation in robot-assisted educational environments. The study highlights the importance of selecting robust perception algorithms when designing interactive educational robotic systems.

## 1 Introduction

Social robots are increasingly being integrated into educational environments to enhance student engagement, collaboration, and learning through natural multimodal interaction. Humanoid platforms such as Pepper have been successfully employed as tutors, teaching assistants, and quiz moderators, demonstrating improvements in learner participation and motivation across diverse educational settings [1, 2, 3]. Their ability to combine speech, gestures, facial expressions, and interactive feedback enables engaging learning experiences that extend beyond conventional computer-based educational systems.

Although educational Human–Robot Interaction (HRI) has advanced considerably, most existing research has focused on social and pedagogical factors that influence learner engagement, including robot appearance, verbal communication, emotional expression, and instructional strategies [1, 4, 5]. Comparatively little attention has been given to the reliability of the robot's underlying perception algorithms. During interactive learning activities, robots continuously interpret sensory information before making decisions that directly affect learners. Incorrect perception may result in unfair interactions, reduced trust, and lower learner confidence, ultimately diminishing students' motivational experiences [6, 7].

Reliable sound event detection is particularly important in robot-mediated educational quiz games, where the robot must accurately identify the first student to respond. Existing sound classification research has primarily evaluated algorithms using engineering metrics such as accuracy, precision, latency, and robustness [8, 9, 10]. Deep learning approaches, especially Convolutional Neural Networks (CNNs), have achieved excellent performance under controlled laboratory conditions but often experience performance degradation when deployed in acoustically different environments because of domain mismatch [11, 12]. In contrast, template-based signal processing methods such as Cross-Correlation compare incoming signals with an environment-specific reference template, providing robust real-time detection for single-event recognition tasks without requiring extensive retraining [9, 13].

Despite significant advances in educational robotics and sound event detection, these research areas have largely evolved independently. Educational HRI studies generally assume reliable robot perception, whereas sound classification research rarely investigates how algorithmic performance influences learners' experiences. Consequently, little empirical evidence exists regarding whether improvements in perception accuracy translate into measurable improvements in student motivation during robot-assisted learning.

To address this gap, this paper investigates the influence of sound detection accuracy on learner motivation using a Pepper robot hosting a robot-mediated educational quiz game. Two sound detection approaches a CNN-based classifier and a Cross-Correlation algorithm were compared in a controlled between-subjects experiment involving 40 university students, with 20 participants assigned to the CNN control group and 20 participants assigned to the Cross-Correlation experimental group. Both groups completed the same quiz under identical conditions, with the sound detection algorithm representing the only experimental variable. Student motivation was evaluated using the Intrinsic Motivation Inventory (IMI), while algorithm performance was assessed through real-time sound detection during gameplay.

Building upon Self-Determination Theory [14, 15], this study introduces the Algorithmic Precision–Motivation Relationship (APMR) model, which proposes that improvements in perception accuracy positively influence learner motivation by enhancing perceptions of fairness, competence, and engagement during robot-mediated interaction. Unlike previous educational HRI studies [16, 17, 18, 19] that primarily investigate social behaviours while assuming technically reliable systems, this work considers the perception algorithm itself as the principal experimental variable.

The main contributions of this paper are threefold:

1. A comparative evaluation of CNN and Cross-Correlation sound detection algorithms for real-time first-responder identification in a Pepper-mediated educational quiz game.
2. An empirical investigation of the relationship between algorithmic accuracy and students' intrinsic motivation using the Intrinsic Motivation Inventory.
3. Empirical validation of the proposed Algorithmic Precision–Motivation Relationship (APMR) model, highlighting perception accuracy as an important design consideration for educational robotic systems.

# 2 Proposed System and Experimental Design

## 2.1 System Overview

To investigate the influence of algorithmic accuracy on learner motivation, an interactive robot-mediated quiz system was developed using the Pepper humanoid robot. The robot acted as the quiz host, presenting multiple-choice questions, detecting the first student to press a physical buzzer, announcing the selected participant, providing immediate verbal and gestural feedback, and maintaining the game scoreboard through its tablet interface. The complete interaction pipeline is illustrated in Figure 1.

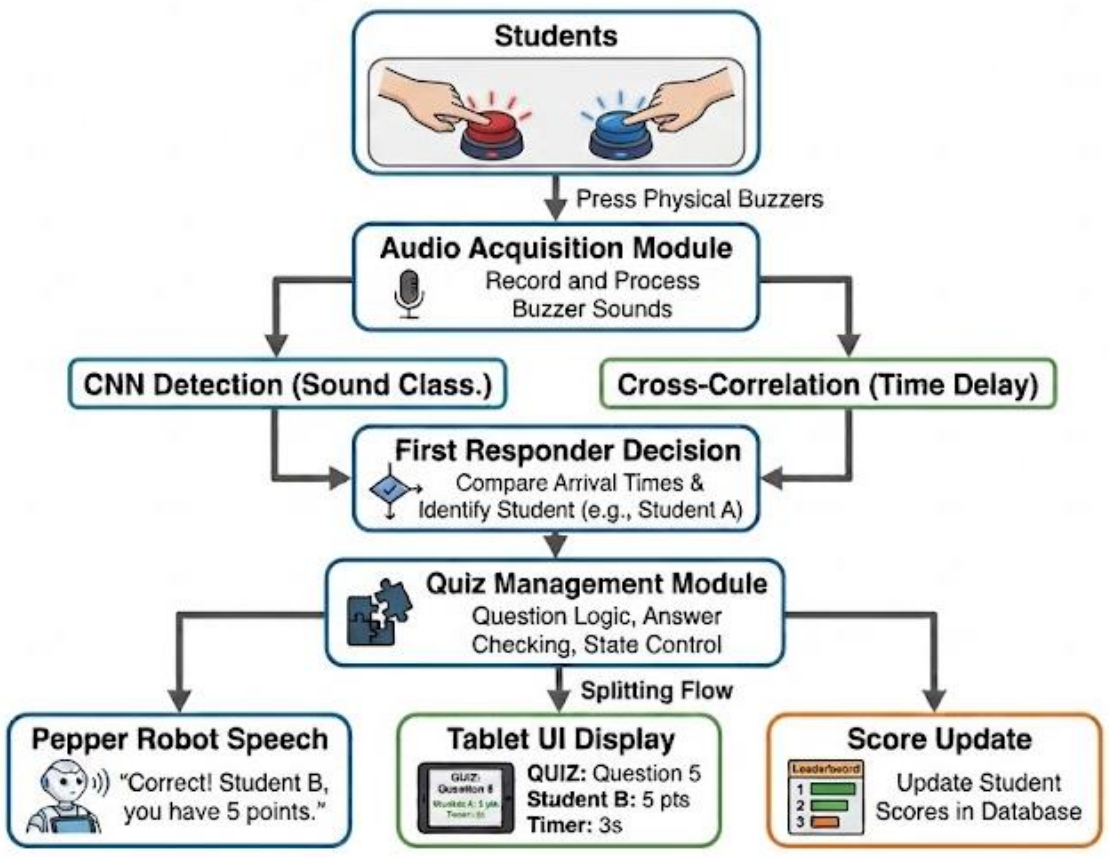

**Fig. 1.** Overall architecture of the proposed Pepper-based quiz system showing the interaction between students, audio acquisition, sound detection, quiz management, and robot feedback modules.

The system consists of four principal components: (1) the Pepper robot, responsible for dialogue management and user interaction; (2) an audio acquisition module that continuously captures the buzzer sounds through a single channel microphone; (3) a sound detection module implementing either a Convolutional Neural Network (CNN) or a Cross-Correlation algorithm; and (4) the quiz management module that controls question presentation, score calculation, and feedback generation. Except for the sound detection algorithm, all other system components remained identical throughout the experiment, ensuring that any observed differences in learner motivation could be attributed solely to algorithmic performance.

The interaction sequence begins when Pepper presents a quiz question. Students respond by pressing their physical buzzers, generating acoustic signals that are processed by the active detection algorithm. The identified first responder is immediately announced by Pepper, after which the student provides an answer. Correct responses increase the player's score, while Pepper provides verbal encouragement and updates the live leaderboard displayed on its tablet.

### 2.2 Sound Detection Algorithms

Two alternative sound detection approaches were evaluated in this study.

The first approach employed a Convolutional Neural Network (CNN) trained to classify buzzer sounds from short audio segments. The network was trained using recordings collected under controlled laboratory conditions and subsequently deployed without additional retraining. During gameplay, incoming audio was continuously analysed and the model predicted whether a buzzer event had occurred based on its learned acoustic features.

The second approach employed a Cross-Correlation algorithm that compared the incoming audio signal with a reference buzzer template recorded immediately before each experimental session. Instead of relying on previously learned representations, this method directly measures waveform similarity, allowing the detector to adapt naturally to the acoustic characteristics of the deployment environment. A detection event was generated when the normalized correlation coefficient exceeded a predefined threshold.

The two algorithms represent contrasting design philosophies. The CNN is a data-driven learning approach that requires representative training data and generally performs well under conditions similar to those encountered during training. In contrast, Cross-Correlation is a template-based signal-processing approach that requires minimal calibration but offers greater robustness when environmental acoustic conditions differ from the original training environment. This comparison provides an opportunity to evaluate not only algorithmic performance but also the educational consequences of deployment robustness.

### 2.3 Experimental Design

A controlled between-subjects experimental design was employed to evaluate the impact of sound detection accuracy on student motivation in a robot-mediated educational quiz. Forty university students voluntarily participated in the study and were equally assigned to one of two experimental conditions. The control group (n = 20) interacted with a Pepper robot using a CNN-based sound detection algorithm, while the experimental group (n = 20) interacted with an identical quiz system employing the proposed Cross-Correlation algorithm.

Each experimental condition was divided into four teams of five students. Two teams competed against each other during each quiz session, resulting in two competitive sessions for the CNN condition and two corresponding sessions for the Cross-Correlation condition. To ensure a fair comparison, both groups completed the same quiz using identical questions, scoring rules, robot behaviours, classroom layout, and interaction procedures. The only experimental variable was the sound detection algorithm used to identify the first student who pressed the buzzer.

Each session began with a brief introduction to the game rules followed by a short practice round. Pepper then autonomously conducted the quiz by presenting questions, detecting the first responder, validating answers, updating team scores, and providing verbal and gestural feedback throughout the activity. At the end of each session, participants completed the Intrinsic Motivation Inventory (IMI) questionnaire individually before the debriefing.

Student motivation was evaluated using five IMI subscales: Interest/Enjoyment, Perceived Competence, Effort/Importance, Perceived Choice, and Pressure/Tension. Each item was rated on a seven-point Likert scale. In parallel, the system automatically logged the detection decisions and algorithm performance throughout each session. Descriptive statistics were performed to compare the motivational outcomes between the CNN and Cross-Correlation conditions. The relationship between algorithmic performance and student motivation was further examined to evaluate the proposed Algorithmic Precision–Motivation Relationship (APMR) model.

## 3 Results and Analysis

This section presents the experimental results obtained from the comparative evaluation of the CNN-based and Cross-Correlation sound detection algorithms. The analysis focuses on two aspects: (1) the effect of the sound detection algorithm on students' intrinsic motivation, measured using the Intrinsic Motivation Inventory (IMI), and (2) the practical performance of the two algorithms during the robot-mediated quiz sessions.

### 3.1 Intrinsic Motivation Assessment

Table 1 summarises the descriptive statistics for the five IMI subscales. For consistency of interpretation, the Pressure/Tension subscale was reverse-coded before analysis; therefore, higher values on all reported subscales represent more positive motivational outcomes.

The experimental group, which interacted with the Cross-Correlation-based system, consistently achieved higher mean scores across all motivational dimensions than the control group using the CNN-based system. The largest improvement was observed for the Interest/Enjoyment subscale, where the mean score increased from 3.49 (SD = 0.56) in the CNN condition to 4.38 (SD = 0.54) in the Cross-Correlation condition. Similarly, Perceived Competence increased from 3.74 (SD = 0.58) to 4.31 (SD = 0.48), suggesting that students felt more confident and capable when the robot accurately identified the first responder.

The Cross-Correlation condition also produced higher scores for Effort/Importance, increasing from 3.80 (SD = 0.38) to 4.60 (SD = 0.46), indicating that students invested greater effort and remained more engaged throughout the quiz activity. Although the difference in Perceived Choice was comparatively smaller, participants interacting with the Cross-Correlation system still reported greater autonomy during gameplay. After reverse coding, the Pressure/Tension subscale likewise demonstrated a more favourable outcome for the experimental group, indicating that students experienced lower psychological pressure and greater comfort when interacting with the more reliable detection system.

Overall, these findings indicate that improvements in algorithmic accuracy positively influenced multiple dimensions of intrinsic motivation, supporting the central hypothesis that reliable robot perception contributes to a more engaging educational experience.

**Table 1.** Descriptive statistics of IMI subscale scores for the CNN and Cross-Correlation conditions.

| IMI Subscale | CNN Mean ± SD | Cross-Correlation Mean ± SD |
|---|---|---|
| Interest/Enjoyment | 3.49 ± 0.56 | 4.38 ± 0.54 |
| Perceived Competence | 3.74 ± 0.58 | 4.31 ± 0.48 |
| Effort/Importance | 3.80 ± 0.38 | 4.60 ± 0.46 |
| Perceived Choice | 4.13 ± 0.61 | 4.27 ± 0.71 |
| Pressure/Tension* | 3.86 ± 0.57 | 4.34 ± 0.63 |

### 3.2 Algorithm Performance Comparison

The performance of the two sound detection algorithms as shown in Table 2 was evaluated during live classroom deployment. The Cross-Correlation algorithm consistently demonstrated superior robustness in recognising the first buzzer presses under realistic acoustic conditions. Unlike the CNN model, which relied on features learned from laboratory training data, the Cross-Correlation approach utilised an environment-specific reference template, allowing it to adapt to classroom acoustics without additional model retraining.

The improved robustness resulted in more reliable first-responder identification and fewer incorrect detections throughout the quiz sessions. Consequently, students experienced fewer interruptions caused by misclassification and perceived the robot's decisions as more consistent and fairer. These observations correspond with the higher motivation scores reported by participants in the experimental condition and provide practical evidence that deployment robustness directly influences the quality of robot-supported educational interaction.

**Table 2.** Comparison of the CNN and Cross-Correlation sound detection algorithms under classroom deployment conditions.

| Performance Metric | CNN | Cross-Correlation |
|---|---|---|
| Training required | Yes | No |
| Calibration | No | Short template recording |
| Detection latency | 300 ms | 280 ms |
| Detection accuracy | 69 % | 87 % |

### 3.3 Summary of Findings

The experimental findings demonstrate that the Cross-Correlation-based detection system provided a consistently better learning experience than the CNN-based approach. Higher detection reliability was associated with increased student interest, greater perceived competence, stronger engagement, and lower perceived pressure during the quiz activity. These results provide empirical support for the proposed Algorithmic Precision–Motivation Relationship (APMR) model, suggesting that algorithmic accuracy should be considered an important design factor when developing robot-assisted educational systems.

## 4 Discussion

The findings of this study demonstrate that the selection of a sound detection algorithm can significantly influence students' motivational experiences during robot-mediated learning. While previous research has primarily attributed learner engagement to the robot's social behaviours, communication strategies, and pedagogical design, the present study shows that the technical reliability of the robot's perception system also plays a critical role. Participants interacting with the Cross-Correlation-based system consistently reported higher levels of interest, perceived competence, effort, perceived choice, and lower perceived pressure (after reverse coding) than those interacting with the CNN-based system.

The superior motivational outcomes observed in the experimental condition can be explained by the greater robustness of the Cross-Correlation algorithm under real classroom conditions. Because the algorithm relied on an environment-specific reference template rather than a pre-trained statistical model, it was less affected by acoustic variability and produced more reliable first-responder detection. Consequently, students experienced fewer incorrect robot decisions, resulting in a more consistent perception of fairness throughout the quiz. These findings support previous Human–Robot Interaction research suggesting that users rapidly lose confidence when robots behave unpredictably or make incorrect decisions. In educational settings, where fairness directly influences learners' willingness to participate, reliable perception appears to be as important as natural interaction behaviours.

The results also provide initial empirical support for the proposed Algorithmic Precision–Motivation Relationship (APMR) model. The model proposes that improvements in algorithmic accuracy enhance learners' motivational experiences by increasing perceived competence, fairness, and engagement during interaction. Rather than treating perception algorithms solely as engineering components, this work demonstrates that they should also be considered educational design variables capable of influencing learning experiences. This perspective extends existing research in educational robotics by connecting technical system performance with established motivational theory.

Although the findings are encouraging, several limitations should be acknowledged. The study involved a relatively small sample of university students and evaluated a single educational activity using one humanoid robot platform. Furthermore, only two sound detection approaches were compared under a specific classroom environment. Future research should investigate larger and more diverse participant populations, additional educational scenarios, and modern deep learning architectures incorporating domain adaptation and continual learning techniques. Examining other perception modalities, including vision-based gesture recognition and multimodal sensor fusion, would further improve understanding of how algorithmic reliability influences human–robot interaction in educational contexts.

## 5 Conclusion

This paper investigated the influence of sound detection accuracy on student motivation in a Pepper robot-mediated educational quiz. A controlled experiment involving forty university students compared a CNN-based sound detection algorithm with a Cross-Correlation approach under identical classroom conditions. The results demonstrated that the Cross-Correlation algorithm produced superior motivational outcomes across all measured IMI dimensions, indicating that robust and reliable perception contributes directly to a more engaging learning experience.

Beyond comparing two sound detection algorithms, this study highlights an important shift in educational Human–Robot Interaction research by demonstrating that algorithmic accuracy is not merely a technical performance metric but also a factor affecting learners' motivation and perceptions of fairness. The proposed Algorithmic Precision–Motivation Relationship (APMR) model provides an initial conceptual framework linking perception accuracy with motivational outcomes in robot-assisted learning.

Future work will extend this research by evaluating additional perception algorithms, larger participant populations, and diverse educational environments while investigating the applicability of the APMR model across different robot platforms and interactive learning activities.

## References

1. Belpaeme, T., Kennedy, J., Ramachandran, A., Scassellati, B., & Tanaka, F. (2018). Social robots for education: A review. Science Robotics, 3(21), eaat5954. https://doi.org/10.1126/scirobotics.aat5954
2. Tekerek, M., Beyazaslan, Z., Aydemir, H. et al. A systematic literature review and mapping of human-robot interaction in educational contexts. Humanit Soc Sci Commun 13, 336 (2026). https://doi.org/10.1057/s41599-026-06698-y
3. Mubin, O., Stevens, C. J., Shahid, S., Al Mahmud, A., & Dong, J.-J. (2013). A review of the applicability of robots in education. Technology for Education and Learning, 1, 1–7. https://doi.org/10.2316/Journal.209.2013.1.209-0015
4. Nasir, J., Bruno, B., & Dillenbourg, P. (2024). Social robots as skilled ignorant peers for supporting learning. Frontiers in Robotics and AI, 11, Article 1385780. https://doi.org/10.3389/frobt.2024.1385780
5. Tutul, R., Buchem, I., & Pinkwart, N. (2025). Design and evaluation of a sound-driven robot quiz system with fair first-responder detection and gamified multimodal feedback. Robotics, 14(9), 123. https://doi.org/10.3390/robotics14090123
6. Alhaji, B., Büttner, S., Sanjay Kumar, S., & Prilla, M. (2025). Trust dynamics in human interaction with an industrial robot. Behaviour & Information Technology, 44(2), 266–288. https://doi.org/10.1080/0144929X.2024.2316284
7. Tozadore DC and Romero RAF (2024) Multiuser design of an architecture for social robots in education: teachers, students, and researchers perspectives. Front. Robot. AI 11:1409671. doi: 10.3389/frobt.2024.1409671
8. Virtanen, T., Plumbley, M. D., & Ellis, D. (Eds.). (2018). Computational analysis of sound scenes and events. Springer. https://doi.org/10.1007/978-3-319-63450-0
9. Hajihashemi, V., Gharahbagh, A. A., Machado, J. J. M., & Tavares, J. M. R. S. (2024). Audio event detection based on cross correlation in selected frequency bands of spectrogram. In A. Rocha et al. (Eds.), Information Systems and Technologies: WorldCIST 2023. Lecture Notes in Networks and Systems, vol. 802. Springer. https://doi.org/10.1007/978-3-031-45651-0_19
10. Chu, H.-C., Zhang, Y.-L., & Chiang, H.-C. (2023). A CNN sound classification mechanism using data augmentation. Sensors, 23(15), 6972. https://doi.org/10.3390/s23156972
11. Mohaimenuzzaman, M., Bergmeir, C., West, I., & Meyer, B. (2023). Environmental sound classification on the edge: A pipeline for deep acoustic networks on extremely resource-constrained devices. Pattern Recognition, 133, 109025. https://doi.org/10.1016/j.patcog.2022.109025
12. Nogueira, A. F. R., Oliveira, H. S., Machado, J. J. M., & Tavares, J. M. R. S. (2022). Sound classification and processing of urban environments: A systematic literature review. Sensors, 22(22), 8608. https://doi.org/10.3390/s22228608
13. Plenkers, K., Ritter, J.R.R. & Schindler, M. Low signal-to-noise event detection based on waveform stacking and cross-correlation: application to a stimulation experiment. J Seismol 17, 27–49 (2013). https://doi.org/10.1007/s10950-012-9284-9
14. Ryan, R. M. (1982). Control and information in the intrapersonal sphere: An extension of cognitive evaluation theory. Journal of Personality and Social Psychology, 43(3), 450–461. https://doi.org/10.1037/0022-3514.43.3.450
15. Ryan, R. M., & Deci, E. L. (2020). Intrinsic and extrinsic motivation from a self-determination theory perspective: Definitions, theory, practices, and future directions. Contemporary Educational Psychology, 61, 101860. https://doi.org/10.1016/j.cedpsych.2020.101860

16. Tutul, R., Buchem, I., Jakob, A., & Pinkwart, N. (2026). Evaluating Intrinsic Motivation in Robot-Supported Quiz-Based Learning: A Comparative Study of Verbal-Only and Multimodal Feedback with Sound Input. Acta Informatica Pragensia, 15(1), Forthcoming article. https://doi.org/10.18267/j.aip.292
17. R. Tutul, I. Buchem, A. Jakob and N. Pinkwart, "Technology Acceptance in University Robot-Supported Quiz-Based Learning: Verbal-Only Versus Multimodal Feedback With Sound Input," in IEEE Access, vol. 14, pp. 956-965, 2026, https://doi.org/10.1109/ACCESS.2025.3648556
18. Buchem, I., Tutul, R., Jakob, A., Pinkwart, N. (2023). Non-verbal Sound Detection by Humanoid Educational Robots in Game-Based Learning. Multiple-Talker Tracking in a Buzzer Quiz Game with the Pepper Robot. RiE 2023. Lecture Notes in Networks and Systems, vol 747. Springer, Cham. https://doi.org/10.1007/978-3-031-38454-7_16
19. Buchem, I., Bonga, G.A.K., Tutul, R. (2024). From Screens to AI-Driven Human-Robot Interaction: Comparing Students' Perceptions of the Robots Pepper and Furhat for ChatGPT-Based Exam Preparation. RiE 2024. Lecture Notes in Networks and Systems, vol 1084. Springer, Cham. https://doi.org/10.1007/978-3-031-67059-6_34